\documentclass[notitlepage,prl,twocolumn,aps, amsmath, amssymb]{revtex4-1}
\usepackage{epsfig}
\usepackage{graphicx}

\newcommand{\vT}{{\mbox{\boldmath$T$}}}

\newcommand{\vk}{{\mbox{\boldmath$k$}}}

\newcommand{\vsk}{{\small \mbox{\boldmath$k$}}}

\newcommand{\vg}{\mbox{\boldmath$g$}}

\newcommand{\vsig}{\mbox{\boldmath$\sigma$}}
\newcommand{\vd}{\mbox{\boldmath$d$}}

\begin{document}

\title{Hexagonal pnictide SrPtAs: superconductivity with locally broken inversion symmetry}

\author{S.J. Youn $^{1,2}$, M.H. Fischer$^{3,4}$, S.H. Rhim$^{1,5}$, M. Sigrist $^4$, and D.F. Agterberg$^5$}

\address{$^1$ Department of Physics and Astronomy, Northwestern University, Evanston, Illinois, 60208-3112, USA}
\address{$^2$ Department of Physics Education and Research Institute of Natural Science, Gyeongsang National University, Jinju 660-701, Korea}
\address{$^3$ LASSP, Department of Physics, Cornell University, Ithaca, NY 14850 USA}
\address{$^4$ Theoretische Physik ETH-H\"onggerberg CH-8093 Z\"urich, Switzerland}
\address{$^5$ Department of Physics, University of Wisconsin-Milwaukee, Milwaukee, WI 53211, USA}


\date{\today}
\begin{abstract}
  Unlike other pnictides, \mbox{SrPtAs} has a hexagonal structure,
  containing layers with As-Pt atoms that form a honeycomb lattice.
  These layers lack inversion symmetry which allows for a spin-orbit coupling
  that we show has a dramatic effect on superconductivity in this material.
  In particular, for conventional $s$-wave superconductivity in SrPtAs,
  both the spin susceptibility and the paramagnetic limiting field are enhanced significantly
  with respect to that usually expected for $s$-wave superconductors.
  SrPtAs provides a prime example of a superconductor with locally broken inversion symmetry.

\end{abstract}
\maketitle

\begin{figure}[b]
\epsfxsize=\hsize  \center{\epsfbox{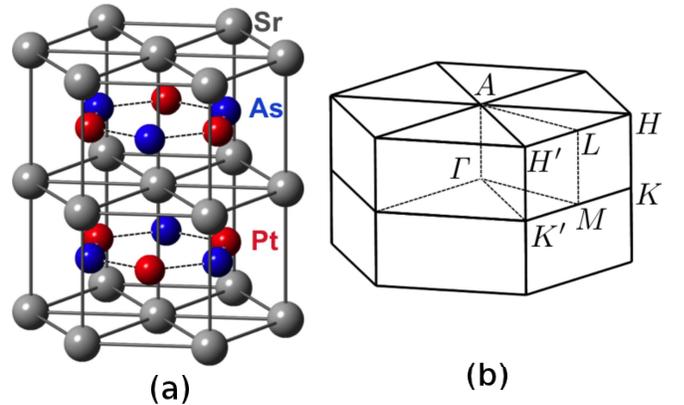}}
\caption{(a) Structure and (b) Brillouin zone in SrPtAs.
  Red, blue, and grey spheres denote Pt, As, and Sr atoms, respectively.
  \label{fig1}}
\label{fig:structure}
\end{figure}

The superconducting pnictides \cite{kam08} present a fascinating class of materials that highlight the interplay between electronic correlations, superconductivity, and magnetism in a multi-orbital system \cite{maz09}. SrPtAs is a new member to this family with a unique feature: the As-Pt atoms in a single layer form a honeycomb lattice \cite{yos11}, see Fig.~\ref{fig:structure}. It is natural to ask if the new lattice structure can have a consequence on superconductivity. Here we argue that it does. In particular, even though SrPtAs has a center of inversion symmetry, the broken inversion symmetry inherent to a single As-Pt layer has non-trivial consequences. Assuming that SrPtAs is a spin-singlet superconductor, we show that it is expected to have a non-vanishing spin susceptibility at zero temperature with a magnitude that is a significant portion of the normal state spin-susceptibility. We further show that it is likely to have a critical field larger than the paramagnetic limiting field.

Of particular importance to this work is that a single As-Pt layer lacks a center of inversion symmetry. The unit cell of SrPtAs  contains  two inequivalent As-Pt layers that are related by inversion symmetry, see Fig.~\ref{fig:structure}. As shown below, SrPtAs also has a spin-orbit coupling (SOC) that is larger than the inter-layer coupling. This combination of broken local inversion and strong spin-orbit coupling allows SrPtAs to be a good candidate for superconductivity with local inversion-symmetry breaking. We use this term to refer to the fact that physical properties usually associated with non-centrosymmetric superconductivity appear in SrPtAs, despite the presence of a center of inversion symmetry. For spin-singlet superconductors, these properties include an enhanced paramagnetic depairing field and a non-vanishing spin-susceptibility at zero temperature \cite{bul76,gor01,fri04,fri04-2,sam05,kim07}. In the following, we initially present the electronic structure of SrPtAs and then turn to an examination of the superconducting state in this material.

\begin{figure}[t]
\epsfxsize=\hsize \center{\epsfbox{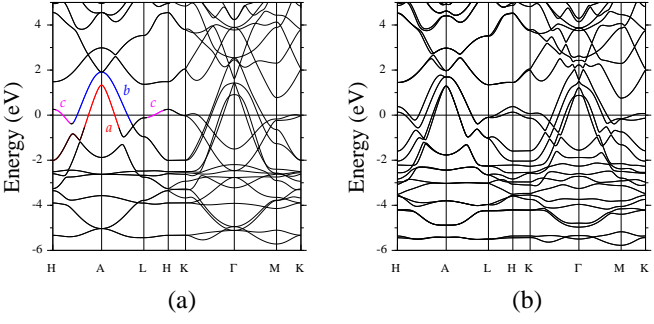}}
\caption{Energy bands of SrPtAs  (a) without and (b) with spin-orbit coupling. Zero energy represents the Fermi level. Indices $a$, $b$, and $c$ in (a) represent the three bands crossing the Fermi level.
\label{fig_band}}
\end{figure}

\begin{figure}[tp]
\epsfxsize=\hsize \center{\epsfbox{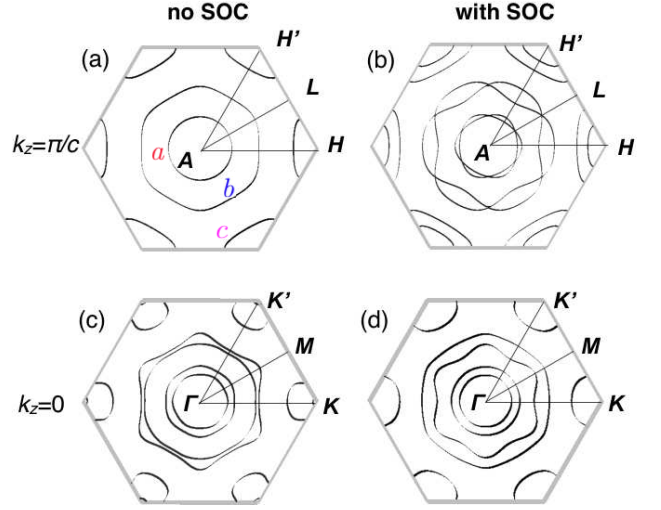}}
\caption{Cross sections of the Fermi surface of SrPtAs  with and without spin-orbit coupling. Figure (a) [(c)] is for $k_z=\pi/c$ and no spin-orbit [with spin-orbit], while Figure (b) [(d)] is for $k_z=0$ and no spin-orbit [with spin-orbit]. Indices $a$, $b$, and $c$ in (a) represent the three bands crossing the Fermi level.\label{fig2}}
\end{figure}

First-principles calculations were performed using the highly precise
full-potential linearized augmented plane wave (FLAPW) method \cite{flapw}.
We have used the experimental lattice constants $a=4.24 \AA$ and $c=8.98 \AA$\cite{wenski86}
and a cutoff of 186 eV for basis functions. The local density approximation (LDA) is
used for the exchange-correlation as parametrized by Hedin and Lundqvist\cite{LDA},
and spin-orbit coupling has been calculated using a second-variational treatment\cite{soc}.
Fig.~\ref{fig_band} and Fig.~\ref{fig2} show the results of LDA calculations with and without spin-orbit coupling.
Energy bands near the Fermi level originate from Pt 5$d$ and As 4$p$ orbitals.
Specifically, the Fermi surface sheets labeled $a$ and $b$ in Fig.~\ref{fig2}
(a) stem from Pt $d_{xy}$, $d_{x^2-y^2}$, As $p_x$, and $p_y$ orbitals
while that labeled $c$ stems from Pt $d_{xz}$, $d_{yz}$ and As $p_z$ orbitals.
Our results without spin-orbit coupling agree with those of Ref.~\cite{she11}. Note the qualitative changes when spin-orbit coupling is added.
In particular, the spin-orbit coupling leads to appreciable changes in the band structure
along the symmetry lines of $H-A$ and $L-H$.
Also of relevance is the difference between the bands along the symmetry lines $H-A-L$ and $K-\Gamma-M$
when there is no spin-orbit coupling (Fig.~\ref{fig_band}).
This difference is due to inter-layer coupling between the As-Pt layers.
This coupling vanishes for symmetry reasons in the plane given by $k_z=\pi/c$.
The band structure reveals that the band splittings due to spin-orbit coupling are
comparable to or larger than those due to inter-layer coupling. A fact that plays an important role in the superconducting state.

To understand the bands stemming from the LDA calculations, it is useful to consider initially a single As-Pt layer. A key point is that this layer does not have a center of inversion and, therefore, a spin-orbit coupling term of the form
\begin{equation}
  \mathcal{H}^i_{so} = \alpha_i\sum_{\vsk,s,s'}
  \vg_{\vsk} \cdot \vsig_{ss'} c_{\vsk si}^{\dag} c^{\phantom{\dag}}_{\vsk s'i}
\label{eq-1}
\end{equation}
exists, where $ c_{\vsk si}^{\dag} $ ($ c^{\phantom{\dag}}_{\vsk si} $) creates (annihilates)
an electron with  momentum $ \vk $  and pseudo-spin $s$ in layer $i$, $\vsig$ denote the Pauli matrices, and $\alpha_i$ is the layer $i$ spin-orbit coupling energy. Time-reversal symmetry imposes $\vg_{\vsk}=-\vg_{-\vsk}$. The invariance of the Hamiltonian under point-group operations leads to the requirement $\vg_{\vsk}=G_s\vg_{G^{-1}\vsk}$,
where $G_s$ is the rotation matrix for a pseudo-vector and $G$ is the rotation matrix for a vector. The point group of a As-Pt layer is $D_{3h}$. Invariance under the mirror symmetries with normals along the $c$ axis and along the Pt-Pt bond imply that $\vg(k_x,0,0)=\vg(k_x,0,\pi/c)=0$ (here a Pt-Pt bond is taken to be along the $y$ axis). This reveals itself for bands along the $A$ to $L$ direction in the Brillouin zone, where there is no spin-orbit splitting, see Fig.~3(b). Within a tight-binding approach, we find $\vg_{\vsk}=\hat{z}\sum_i \sin(\vsk\cdot \vT_i)$, where $\vT_i$ are the translation vectors $\vT_1=(0,a,0)$, $\vT_2=(\sqrt{3}a/2,-a/2,0)$, and $\vT_3=(-\sqrt{3}a/2,-a/2, 0)$.
This form of spin-orbit coupling can be found for all bands stemming from Pt $d$ orbitals by including hopping to neighboring As $p$ orbitals  and by including on-site ${\bf L}\cdot {\bf S}$ for both As and Pt sites. We note that this spin-orbit coupling is similar to that used by Kane and Mele to discuss the quantum spin hall effect in graphene \cite{kan05}. Symmetry also allows for $g_x$ and $g_y$ to be non-zero. However these components must be odd in $k_z$ and, within a tight-binding analysis, are only found by including hopping along the $c$-axis. Given the much weaker dispersion of the bands along $k_z$ relative to the in-plane dispersion, we expect that $g_x$ and $g_y$ are much smaller than $g_z$ and we will only include $g_z$ in the following.

The analysis in the previous paragraph applies to a single As-Pt layer. The two inequivalent As-Pt layers are related by inversion symmetry, consequently, $\alpha_i$ is of opposite sign for the two layers, i.e., $\alpha_i = (-1)^i \alpha$. To complete the description for the solid, a coupling between the two inequivalent layers is required. We take this to be $\epsilon_c(\vsk)$ (symmetry requires this to vanish for $k_z=\pi/c$).
Provided there are no band degeneracies other than spin and layer degeneracies, a generic Hamiltonian for SrPtAs is then
\begin{multline}
  \mathcal {H}_0 = \sum_{\vsk} \Psi^{\dagger}(\vsk)\Big\{ [\epsilon_1({\vsk})-\mu] \sigma_0\tau_0 + \alpha(\vsk)
\sigma_z\tau_z\\
+Re[\epsilon_c(\vsk)]\sigma_0\tau_x+ Im[\epsilon_c(\vsk)]\sigma_0\tau_y\Big\}\Psi({\vsk}),
\label{eq-2}
\end{multline}
where $\Psi({\vsk})=(c_{\vsk \uparrow 1},c_{\vsk \downarrow 1},c_{\vsk \uparrow 2},c_{\vsk \downarrow 2})^T$, $\sigma_i$ ($\tau_i$) are Pauli matrices that operate on the pseudo-spin (layer) space and $\alpha(\vsk)=\alpha g_z(\vsk)$.
This Hamiltonian can be diagonalized with resulting dispersion relations $\epsilon_\pm({\bf k})=\epsilon_1({\bf k})\pm\sqrt{|\epsilon_c({\bf k})|^2+\alpha^2({\bf k})}$ and each state has a 2-fold Kramers degeneracy. To gain an intuition for the terms appearing in this Hamiltonian, we state the results for a simple tight-binding theory (note that below we keep these terms arbitrary). This yields $\epsilon_1({\bf k})=t_1(\cos\vk\cdot {\bf T_1}+\cos\vk\cdot {\bf T_2}+\cos\vk\cdot {\bf T_3})+t_{c2}\cos(c k_z)$ and $\epsilon_c({\bf k})= t_c \cos(k_zc/2)(1+e^{-i{\bf k}\cdot {\bf T}_3}+e^{i{\bf k}\cdot {\bf T}_2})$ (note that $\vg({\vsk})$ and ${\bf T_i}$ are given above).

From the point of view of superconductivity, the strong spin-orbit coupling is of interest.  In the limit that the inter-layer coupling vanishes, we have two uncoupled non-centrosymmetric systems. It is known that in non-centrosymmetric spin-singlet superconductors the spin-susceptibility and the paramagnetic limiting field are significantly enhanced, if the spin-orbit coupling strength is much larger than the superconducting gap \cite{bul76,gor01,fri04,fri04-2,sam05,kim07}.  Given the large spin-orbit coupling relative to the inter-layer coupling, it is conceivable that
the behavior of superconducting SrPtAs resembles that of a non-centrosymmetric material.
For this reason we calculate both the spin-susceptibility and the limiting field assuming that superconductivity in SrPtAs is spin-singlet (this is a reasonable assumption comparing with other pnictides superconductors \cite{maz09}). To be concrete we assume intra-layer $s$-wave pairing with an interaction \begin{equation}
\mathcal{H}_{sc}=-V\sum_{k,k',i,s,s'}c^{\dagger}_{\vsk s i}c^{\dagger}_{-\vsk s' i}c_{-\vsk' s' i}c_{\vsk' s i}. \label{Hsc} \end{equation} Note that our results do not depend qualitatively on this choice.

For a system described by the Hamiltonian \eqref{eq-2} and \eqref{Hsc}, the susceptibility in the superconducting and normal state can be calculated using\cite{abr62}
\begin{multline}
  \chi^s_{ij} = -\mu_B^2 T \sum_n\sum_{\vsk}{\rm tr}[\sigma_iG(\vsk, \omega_n)\sigma_{j}G(\vsk, \omega_n) \\
  - \sigma_iF(\vsk,\omega_n)\sigma_j^TF^\dag(\vsk, \omega_n)]
  \label{eq:chi}
\end{multline}
with $G(\vsk, \omega_n)$ and $F(\vsk, \omega_n)$ the normal and anomalous Green's functions in the Matsubara formulation. Note that even for this `one-band' formulation, the Green's functions are $4\times4$ matrices, so that the trace runs over both, layer and spin index. In the notation of the Hamiltonian \eqref{eq-2} there are essentially three bands crossing the Fermi energy in SrPtAs [labeled $a$, $b$, and $c$ in  Figs.~\ref{fig_band}(a)and \ref{fig2}(a)] and we can generalize the above expression to
\begin{equation}
  \chi_{ij}=\sum_{\nu}\chi_{ij}(\nu)
  \label{eq:chi3band}
\end{equation}
where the sum runs over the three bands $\nu=a,b,c$. Below we calculate the susceptibility separately for each band using Eq.~\eqref{eq:chi}.

\begin{figure}
\epsfxsize=\hsize \center{\epsfbox{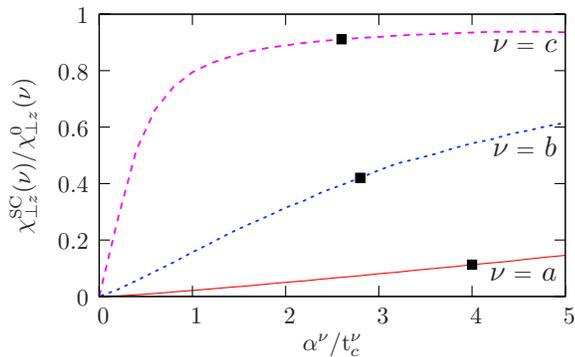}}
\caption{Spin susceptibility at $T=0$ in the superconducting state normalized with respect to the normal state susceptibility for the three bands crossing the Fermi surface as a function of $\alpha^\nu/t^\nu_c$. The black squares denote approximate values for these three bands. \label{fig:susc}}
\end{figure}

In the normal state, $F^\nu(\vsk, \omega_n)=0$, and we find for fields parallel to $z$,
\begin{equation}
  \chi^{0}_{z}(\nu)=2\mu_{B}^{2}\sum_{\vsk,i=\pm}\frac{\partial n_F(\epsilon^\nu_{i}(\vsk))}{\partial \epsilon^\nu_i} = \sum_{\vsk}\chi^{0}_{P}(\vsk, \nu),
  \label{eq:chiznormal}
\end{equation}
where $n_F(\epsilon)$ is the Fermi distribution function as a function of energy $\epsilon$ and $\chi^0_p(\vsk, \nu)$ denotes a Pauli susceptibility for band $\nu$. For fields in plane, we find
\begin{equation}
  \chi^0_{\perp z}(\nu) = \sum_{\vk}\Big\{\frac{|\epsilon^\nu_c(\vsk)|^2\chi^{0}_{P}(\vk, \nu)+[\alpha^\nu(\vsk)]^2\chi^{0}_{vV}(\vk, \nu)}{|\epsilon^\nu_c(\vsk)|^2 + [\alpha^\nu(\vsk)]^2} \Big\},
  \label{eq:chixnormal}
\end{equation}
with the van Vleck susceptibility
\begin{equation}
  \chi_{vV}^0(\vsk, \nu) = 2\mu_{B}^2\Big\{\frac{n_F(\epsilon^\nu_{+}(\vsk))-n_F(\epsilon^\nu_{-}(\vsk))}{\sqrt{|\epsilon^\nu_c(\vsk)|^2 + [\alpha^\nu(\vsk)]^2}}\Big\}.
  \label{eq:vanvleck}
\end{equation}
The Pauli susceptibility describes intra-band processes and at low temperatures is
proportional to the density of states at the Fermi level.
The van Vleck susceptibility describes inter-band processes.
For the superconducting states in the limit $\sqrt{|\epsilon^\nu_c(\vsk)|^2 + [\alpha^\nu(\vsk)]^2}\gg\Delta^\nu$,
we recover the expressions given in Eqs.~\eqref{eq:chiznormal} and \eqref{eq:chixnormal},
where for the Pauli susceptibility, we have to replace $\epsilon^\nu_{\pm}(\vsk)$ with $E^\nu_{\pm}(\vsk) = \sqrt{[\epsilon^\nu_{\pm}(\vsk)]^2+ [\Delta^\nu]^2}$.
The Pauli susceptibility contribution thus vanishes due to the opening of the superconducting gap,
while the van Vleck susceptibility is unchanged by superconductivity, even at $T=0$. Consequently, $\chi^{SC}_z$ will behave like that expected for a conventional spin-singlet superconductor while $\chi^{SC}_{\perp z}$ will have a large spin susceptibility, even at $T=0$. To demonstrate this, Fig.~\ref{fig:susc} shows the ratio of $\chi^{SC}_{\perp z}(\nu)$ in the superconducting phase at $T=0$ to the normal state in-plane susceptibility $\chi_{\perp z}^0(\nu)$ as a function of $\alpha^\nu/t^\nu_c$ for the three bands $\nu=a,b,c$ (where $\alpha^\nu$ is the spin-orbit strength and $t^\nu_c$ is the interlayer coupling strength). These values were determined using simple tight-binding calculations for the three Fermi surface sheets $a,b,c$.
Estimating the ratios $\alpha^\nu/t^\nu_c$ from the band structure,
we find values of $\chi^{SC}_{\perp z}(\nu)/ \chi^{0}_{\perp z}(\nu)=0.11, 0.42$, and $0.91$ for Fermi surface sheet $\nu=a$, $b$, and $c$, respectively.
Consequently, we expect that a sizable portion of the normal state susceptibility will exist in the limit $T\rightarrow 0$ for in-plane magnetic fields. We note that related behavior has recently been predicted for the spin-susceptibility in an examination of the crossover from  non-centrosymmetric to centrosymmetric superconductivity in multi-layer systems \cite{mar11}.

The enhanced susceptibility suggests that the Pauli limiting field will also be enhanced when the field in the basal plane. To calculate this, we include the Zeeman field $\mathcal{H}_Z=\sum_{\vsk,s,s',i} g\mu_B {\bf H}\cdot \vsig_{s,s'}c^{\dagger}_{\vsk s i}c_{\vsk s'i}$\, and orient the field in the basal plane.
Within weak-coupling theory and assuming $\sqrt{|\epsilon_c(\vsk)|^2+\alpha(\vsk)^2}\gg g\mu_B H$ (which is well supported by LDA results), we find the following expression for $T_c$ as a function of $h=g\mu_B|{\bf H}|$:
\begin{equation}
\ln \Big(\frac{T_c}{T_{c0}}\Big) = -\Psi\Big (\frac{1}{2}\Big) + Re\Big\{\frac{1}{2}\Big \langle \Psi\Big(\frac{1}{2}+ih(\vsk)\Big )\Big \rangle_{\vsk}\Big\}, \label{tch}
\end{equation}
where \begin{equation} h(\vsk)=\frac{\frac{h|\epsilon_c(\vsk)|}{2\pi k_B T_c}}{\sqrt{|\epsilon_c(\vsk)|^2+\alpha^2(\vsk)}},\end{equation}
$T_{c0}$ is the transition temperature for $h=0$, $\langle f \rangle_{\vsk}$ means an average of $f$ over the Fermi surface, $Re$ means real part, and $\Psi(x)$ is the digamma function. The band index, $\nu$, is omitted for brevity.
In the limit that $\epsilon_c=0$, we find that $T_c$ is independent of $h$.  This is in agreement with the result predicted and observed for non-centrosymmetric superconductors in the limit of large spin-orbit coupling \cite{fri04,kim07}. Also, in the limit that $\alpha=0$, we find the usual expression for the Pauli limiting field.  At $T_c=0$, using Eq.~\eqref{tch},  we find that the Pauli limiting field is given by
\begin{equation}
\Psi\Big(\frac{1}{2}\Big)=\Big\langle \ln|\frac{h|\epsilon_c(\vsk)|}{2\pi k_B T_{c0}\sqrt{|\epsilon_c(\vsk)|^2+\alpha^2(\vsk)}}|\Big
\rangle_{\vsk}.
\end{equation}
Using tight-binding calculations,
we estimate that the enhancement of the Pauli limiting field,
$h_P/h_{P0}$ (where $h_{P0}$ is the limiting field when $\alpha=0$),
takes the values 1.1, 1.8, and 7.4 for Fermi sheet $a$, $b$, and $c$, respectively.
Provided that the orbital upper critical field is sufficiently large, it should be possible to observe an enhanced Pauli limiting field. For fields along the $c$-axis, a usual Pauli suppression is expected.

Finally, we point out that in addition to spin-singlet pairing, a spin-triplet component will also appear\cite{fis11}. In particular, in a given layer a spin-triplet component with $\vd(\vsk)$ along the direction of $\vg(\vsk)$ exists, such that it has opposite sign in the two inequivalent layers of SrPtAs.

In conclusion, we have shown that the unique structure in the pnictide SrPtAs has non-trivial effects on superconductivity. In particular, the lack of an inversion center in the As-Pt honeycomb lattice layers, combined with strong spin-orbit coupling, allows a significant enhancement of
both the Pauli limiting field and the spin susceptibility for spin-singlet superconductivity. SrPtAs provides an ideal example of superconductivity with locally broken inversion symmetry.

We thank to Michael Weinert, Youichi Yanase, and Daisuke Maruyama for useful discussions.
MHF acknowledges support from NSF Grant No. DMR-0520404 to the Cornell Center for Materials Research and from NSF Grant No. DMR-0955822.
SHR is supported by Department of Energy (DE-FG02-88ER45382).
DFA is supported by NSF grant DMR-0906655. MS is grateful for financial support by the Swiss Nationalfonds and the NCCR MaNEP.


\begin{references}
\bibitem{kam08} Y. Kamihara, T. Watanabe, M. Hirano and H. Hosono, J. Am. Chem. Soc. {\bf 130}, 3296 (2008).
\bibitem{maz09} I.I. Mazin and J. Schmalian, Physica C {\bf 469}, 614 (2009).
\bibitem{yos11} Y. Nishikubo, K. Kudo, M. Nohara, J. Phys. Soc. Jpn {\bf 80}, 055002 (2011).
\bibitem{bul76} L.N. Bulaevskii, A.A. Guseinov, and A.I. Rusinov, Sov. Phys. JETP {\bf 44}, 1243 (1976).
\bibitem{gor01} L.P. Gor'kov and E.I. Rashba, Phys. Rev. Lett. {\bf 87}, 037004 (2001).
\bibitem{fri04} P.A. Frigeri, D.F. Agterberg, A. Koga, and M. Sigrist, Phys. Rev. Lett. {\bf 92}, 097001 (2004).
\bibitem{fri04-2} P.A. Frigeri, D.F. Agterberg, and M. Sigrist, New J. Phys. {\bf 6}, 115 (2004).
\bibitem{sam05} K.V. Samokhin, Phys. Rev. Lett. {\bf 94}, 027004 (2005).
\bibitem{kim07} N. Kimura, K. Ito, H. Aoki, S. Uji, and T. Terashima, Phys. Rev. Lett. {\bf 98}, 197001 (2007).
\bibitem{flapw} E. Wimmer, H. Krakauer, M. Weinert and A. J. Freeman, Phys. Rev. B {\bf 24}, 864 (1981).
\bibitem{wenski86} G. Wenski and A. Mewis, Z. Anorg. Allg. Chem. {\bf 535}, 110 (1986).
\bibitem{LDA} L. Hedin and B. I. Lundqvist, J. Phys. C {\bf 4}, 2064 (1971).
\bibitem{soc} A. H. MacDonald, W. E. Pickett, and D. D. Koelling, J. Phys. C {\bf 13}, 2675 (1980).
\bibitem{she11} I. R. Shein and A. L. Ivanovskii, Physica C {\bf 471}, 594 (2011).
\bibitem{abr62} A.A. Abrikosov and L.P. Gor'kov, Sov. Phys. JETP {\bf 15} 752 (1962).
\bibitem{kan05} C.L. Kane and E.J. Mele, Phys. Rev. Lett. {\bf 95}, 226801 (2005).
\bibitem{mar11} D. Maryama, M. Sigrist,and Y. Yanase, arxiv:1111.4293.
\bibitem{fis11} M.H. Fischer, F. Loder, and M. Sigrist, arxiv:1108.4694.






\end{references}
\end{document}